\documentclass[english,aps,pra,showpacs,preprint]{revtex4}
\usepackage[T1]{fontenc}
\usepackage{color}
\usepackage{babel}
\usepackage{float}
\usepackage{bm}
\usepackage{graphicx}
\usepackage{esint}
\usepackage[unicode=true,pdfusetitle,
 bookmarks=true,bookmarksnumbered=false,bookmarksopen=false,
 breaklinks=false,pdfborder={0 0 0},backref=false,colorlinks=true]
 {hyperref}
\hypersetup{
 citecolor=blue}

\makeatletter
\@ifundefined{textcolor}{}
{%
 \definecolor{BLACK}{gray}{0}
 \definecolor{WHITE}{gray}{1}
 \definecolor{RED}{rgb}{1,0,0}
 \definecolor{GREEN}{rgb}{0,1,0}
 \definecolor{BLUE}{rgb}{0,0,1}
 \definecolor{CYAN}{cmyk}{1,0,0,0}
 \definecolor{MAGENTA}{cmyk}{0,1,0,0}
 \definecolor{YELLOW}{cmyk}{0,0,1,0}
}

\makeatother

\begin{document}

\title{Robust Gaussian Teleportation with Attenuations and Non-unity Gain}

\author{Alencar J. de Faria}

\email{alencar.faria@unifal-mg.edu.br}

\affiliation{Instituto de Ciência e Tecnologia, Universidade Federal de Alfenas,
CEP 37715-400, Poços de Caldas, MG, Brazil}
\begin{abstract}
The average fidelity of the teleportation of a coherent state is calculated
for general Gaussian bipartite systems shared by the partners of the
protocol, Alice and Bob. It is considered that the shared Gaussian
bipartite modes suffer independent attenuations before the processing
of Alice and Bob. Moreover the classical communication between the
partners can be controlled by a gain not necessarily unitary. Comparing
with the classical fidelity threshold of measure-and-prepare methods,
we establish several genuinely quantum teleportation conditions which
depend on the gain and the local attenuations. Considering that the
gain can be tuned to maximize the bipartite state set able to genuinely
quantum teleportation, a condition for teleportation robust to local
attenuations is found. This condition is demonstrated to be essentially
equivalent to the condition of robust Gaussian bipartite entanglement,
obtained in previous articles, showing that the attenuation robustness
is an entanglement property relevant for characterization and application
of bipartite systems. For the derivation of the robust teleportation
conditions, the Gaussian operations onto the bipartite system are
thoroughly studied, so that the transformations that maintain the
fidelity invariant are found. Some scenarios for different Gaussian
bipartite states are presented and discussed.
\end{abstract}


\maketitle

\section{Introduction}

Teleportation was one of the first proposed protocols on quantum information
and it is understood as an elementary piece for more complex quantum
processing and communication. Moreover, it is a resource for the understanding
of fundamental issues, such as the EPR (Einstein-Podolsky-Rosen) paradox
and nonlocality \cite{Bennett93}. There are many reviews available
in literature, being this research field very wide and active \cite{Pirandola06,Furusawa07,Pirandola15,Furusawa-van Loock}.
In particular, implementations of the teleportation over long distances
have advanced and enlarged the scientific and technological frontier
\cite{Landry07,Yin12,Ma12,Herbst15,Takesue15,Wang15}. To perform
the quantum teleportation, two communication stations in different
locations, usually called Alice and Bob, share a bipartite entangled
system, each one retaining a part. Then Alice combines her entangled
subsystem with a signal, without knowing or directly measuring it.
So Alice measures the EPR observables of the combined system and transmits
the obtained classical information to Bob. With the classical instructions,
Bob operates unitarily his entangled subsystem to restore the original
signal. Originally, the teleportation was proposed for discrete variables
systems, posteriorly devised in continuous variables \cite{Vaidman94,Braunstein98}.
The first theoretical and experimental setup for continuous variable
teleportation, and also the most studied, was proposed by Braunstein
and Kimble (BK protocol), using phase and amplitude quadrature optical
modes \cite{Braunstein98,Furusawa98,Bowen03a,Zhang03}. In this context,
the quantum teleportation requires that the bipartite entangled system
should be in an optical two-mode squeezed state. However an ideal
quantum teleportation, with ideal squeezed beams, is physically infeasible,
because the squeezing needs to have infinity rate. Thus, since early
works, the continuous variable teleportation has been studied in realistic
situations, such as finite squeezing, entangled modes subject to lossy
channels \cite{Hofmann01,Chizhov02}, and non-unity gains of the classical
communication \cite{Braunstein01,Ide02,Bowen03b,Namiki08,Namiki11,Chiribella13}.

Thus, even in early teleportation studies, the general connections
between the capacity to perform teleportation and the entanglement
of the quantum resources have been pursued and gradually clarified.
Popescu has already shown that states able to teleportation are not
equivalent to states that violate Bell inequalities \cite{Popescu94}.
Braunstein et al. \cite{Braunstein01} have also shown that the entanglement
for fully symmetric Gaussian bipartite systems is equivalent to the
quantum regime for teleportation performed by such states. However,
in that same article, the authors have presented situations in which
this equivalence are not applicable to non-unity gain teleportation.
After that, Fiurásek \cite{Fiurasek02} have derived the fidelity
of a teleportation performed by a general Gaussian bipartite system,
so that we can notice that the quantum teleportation regime is very
different of the necessary and sufficient entanglement condition,
considering generic shared bipartite systems. On the other hand, considering
local operations to symmetrize and optimize a bipartite system, the
equivalence between Gaussian entanglement of the bipartite system
and its ability to perform teleportation with optimal fidelity above
classical threshold was established by Adesso and Illuminati \cite{Adesso05}.
Therefore, we can notice a complex and deep connection between bipartite
entanglement and its ability to perform quantum teleportation. Realistic
factors, such as limited squeezing, lossy quantum channels and non-unity
gain of the classical communication, make this relation more diverse.
In fact, given that various recent achievements \cite{Landry07,Yin12,Ma12,Herbst15,Takesue15,Wang15}
and proposals \cite{Scheidl13,Bourgoin14,Vallone15,Hosseinidehaj15}
on quantum communication involve long distances, we must consider
that the shared system for teleportation can be asymmetric and suffers
losses. In addition, the effective optimization by local operations
can be unfeasible, because the environmental influences and the long
distance between Alice and Bob make the bipartite states not fully
known. So the present paper is dedicated to detail the conditions
that a general Gaussian bipartite system is able to accomplish quantum
teleportation, considering the afore mentioned realistic factors.
Hence, considering the gain adjustment, the fidelity can be optimized
so that the teleportation is maintained in a quantum regime, robust
to any local partial attenuation. The condition to such robust quantum
teleportation is found and it is shown that this condition is essentially
equivalent to the robust bipartite entanglement condition, found in
previous articles \cite{Barbosa10,Barbosa11}.

To assess the quality of the teleportation, we must use a well-known
quantity, called fidelity \cite{Jozsa94}. So a natural question is
how much fidelity is necessary and sufficient to characterize a genuinely
quantum teleportation, that is, a protocol accomplished exclusively
by quantum processes. Along the years, many different conditions of
quantum teleportation have been developed for different situations.
In the usual continuous variable teleportation proposed by Braunstein
and Kimble, a classical measure-and-prepare strategy, simulating the
same task, has a maximum classical fidelity threshold (CFT) given
by $F_{\mathrm{CFT}}=1/2$ \cite{Braunstein00,Hammerer05}. To establish
the benchmark between classical and quantum regimes, originally the
teleportation was restricted to transmit input coherent states and
using a unity-gain classical communication. However, more recent studies
generalize the benchmark for quantum teleportation to the case of
squeezed state signals \cite{Takei05,Yonezawa07,Chiribella14}. On
the other hand, other studies have pursued the case of non-unity gain
classical communication \cite{Bowen03b,Namiki08,Chiribella13}. In
that case, it is possible to show that gain tuning can improve the
fidelity of the sent signal \cite{Ide02}. Moreover, the teleportation
of the single-photon qubit using hybrid continuous variable schemes
\cite{Takeda13a,Takeda13b,Kogias14} and others variations of the
basic protocol \cite{Neergaard-Nielsen13,He15} that use non-unity
gain have been analyzed theoretical and experimentally. Keeping in
the task of teleporting coherent signals, one can consider a slight
change of the BK teleportation protocol, in which the Alice's input
states set, $\{|\alpha\rangle\}_{A}$, is sent to Bob, so that he
recovers an output states set, ideally represented by $\{|g\alpha\rangle\}_{B}$.
In other words, the signal is teleported and simultaneously amplified
by a gain $g$. Such variant is a generalization of the BK teleportation,
and it is also called teleamplification. 

In order to clarify the correlation properties of Gaussian bipartite
systems, in Section II, it is calculated the average fidelity of the
BK teleportation of an uniform set of coherent states, considering
independent attenuations in quantum channels and non-unity gain of
the classical communication between Alice and Bob. In Section III.
Comparing the calculated fidelity to the CFT with non-unity gain,
we can find a general condition for a Gaussian bipartite system to
perform genuinely quantum teleportation. From symmetry considerations,
we have found other interesting teleportation conditions, since they
do not depend on a full system characterization, despite not being
both necessary and sufficient. From one of these conditions, we show
the gain can be optimized so that the state set of the bipartite system
able to genuine quantum teleportation is maximized. Thus we can establish
a condition to a genuinely quantum teleportation robust to attenuations.
Comparisons between the teleportation conditions and early entanglement
conditions are presented in Section IV, so that the robust quantum
teleportation condition is verified to be essentially equivalent to
the robust entanglement condition \cite{Barbosa10,Barbosa11,Buono12}.
Hence we establish another connection between bipartite entanglement
and teleportation. Some particular cases are studied in Section V,
where the relations among the teleportation and entanglement conditions
are graphically represented. The dynamic of the average fidelity in
terms of the attenuations are plotted in some figures. We compare
symmetric and asymmetric cases, considering the effect of the gain
adjustment. The results are discussed in Section VI.

\section{Non-unity gain teleportation by lossy channels}

To study the teleportation, we should consider a system composed of
three subsystems: the input signal and the pair of correlated modes
shared between Alice and Bob. In the case of continuous variable systems,
we can use the formalism of Wigner distributions, such that the Wigner
function, $W(\bm{x})$, depends on ordinary variables, placed in vector
form as 
\begin{equation}
\bm{x}=(q_{1},p_{1},q_{2},p_{2},...q_{N},p_{N})^{T},\label{vec-scalar}
\end{equation}
in the case of $N$ subsystems. These variables are associated with
the respective quantum operators, 
\begin{equation}
\hat{\bm{x}}=(\hat{q}_{1},\hat{p}_{1},\hat{q}_{2},\hat{p}_{2},...\hat{q}_{N},\hat{p}_{N})^{T},\label{vec-operator}
\end{equation}
that obey the usual commutation relations, $[\hat{q}_{i},\hat{p}_{j}]=2i\delta_{ij}$
and $[\hat{q}_{i},\hat{q}_{j}]=[\hat{p}_{i},\hat{p}_{j}]=0$ \cite{Braunstein05}.
For a physical state represented by a complete density matrix $\rho$,
the mean value of an operator $\hat{\mathcal{O}}$ is calculated by
$\langle\hat{\mathcal{O}}\rangle=\mathrm{tr}(\hat{\mathcal{O}}\rho)$.
So the mean value of the operator $\hat{\bm{x}}$ is 
\begin{equation}
\bm{\mu}=\langle\hat{\bm{x}}\rangle.\label{vec-average}
\end{equation}
We restrict this paper to the case of Gaussian states, so that the
respective Wigner function takes the general form: 
\begin{equation}
W(\bm{x})=\frac{1}{(2\pi)^{2N}\sqrt{\det\bm{V}}}\exp\left[-\frac{1}{2}(\bm{x}-\bm{\mu})^{T}\bm{V}^{-1}(\bm{x}-\bm{\mu})\right],\label{Wigner-function}
\end{equation}
where $\bm{V}$ is the covariance matrix of the complete system, whose
entries are $V_{ij}=\langle\frac{1}{2}\{\Delta\hat{x}_{i},\Delta\hat{x}_{j}\}\rangle$,
such that $\Delta\hat{x}_{i}=\hat{x}_{i}-\mu_{i}$ \cite{Weedbrook12}.

In the teleportation protocol proposed by Braunstein and Kimble, the
signal sent by Alice belongs to a set of pure coherent states, $\{|\alpha\rangle\}_{A}$,
in which the covariance matrix of the set elements is 
\begin{equation}
D=\left(\begin{array}{cc}
1 & 0\\
0 & 1
\end{array}\right),\label{matrix-D}
\end{equation}
and their mean values are $\langle\hat{q}_{S}\rangle=2\Re(\alpha)$
and $\langle\hat{p}_{S}\rangle=2\Im(\alpha)$. Assuming that the input
signal states follow a central Gaussian distribution, 
\begin{equation}
P(\alpha)=\frac{\lambda}{\pi}\exp(-\lambda|\alpha|^{2}),\label{distribution-function}
\end{equation}
so that we obtain an uniform distribution of coherent states taking
$\lambda\rightarrow0$.

Before signal transmission, Alice and Bob have to share each mode
of a bipartite system, which is originally generated as a Gaussian
state, whose covariance matrix is formed by 2x2 matrix blocks in the
following way 
\begin{equation}
V=\left(\begin{array}{cc}
A & C\\
C^{T} & B
\end{array}\right),\label{matrix-V}
\end{equation}
where 
\begin{equation}
A=\left(\begin{array}{cc}
Q_{A} & K_{A}\\
K_{A} & P_{A}
\end{array}\right)\label{matrix-A}
\end{equation}
is the covariance matrix of the mode designed to Alice, 
\begin{equation}
B=\left(\begin{array}{cc}
Q_{B} & K_{B}\\
K_{B} & P_{B}
\end{array}\right)\label{matrix-B}
\end{equation}
is the covariance matrix of the mode designed to Bob, and 
\begin{equation}
C=\left(\begin{array}{cc}
K_{Q} & K_{1}\\
K_{2} & K_{P}
\end{array}\right)\label{matrix-C}
\end{equation}
is the correlation matrix between such subsystems. For simplicity
without losing generality, we consider that the shared bipartite system
has vanishing mean-value canonical operators, because, otherwise,
to restore the teleported signal, Bob only needs to modulate trivial
signal displacements in phase space for compensation of the bipartite
system contribution \cite{Pirandola06}. On the other hand the bipartite
system modes arrive at Alice and Bob after passing through attenuation
channels, which alter them according to \cite{Eisert05} 
\begin{equation}
V_{t}=\mathcal{L}(V)=L(V-I)L+I\label{attenuation}
\end{equation}
such that $V_{t}$ is the covariance matrix of the attenuated bipartite
system and 
\begin{equation}
L=\mathrm{diag}(t_{A},t_{A},t_{B},t_{B}),\label{attenuation-matrix}
\end{equation}
where $t_{A}$ and $t_{B}$ are the channel transmissibilities of
Alice and Bob, respectively. The submatrices of $V_{t}$ are transformed
by $A_{t}=L_{A}(A-I^{(2)})L_{A}+I^{(2)}$, $B_{t}=L_{B}(B-I^{(2)})L_{B}+I^{(2)}$,
and $C_{t}=L_{A}CL_{B}$, with $L_{i}=\mathrm{diag}(t_{i},t_{i})$,
$i=\{A;B\}$ and $I^{(2)}$ is the 2x2 identity matrix.

Since each communication station is in possession of its subsystem
bipartite, Alice combines the input signal with her bipartite subsystem
by a beam-splitter operation and she measures the quadratures $\hat{q}_{-}=(\hat{q}_{A}-\hat{q}_{\mathrm{in}})/\sqrt{2}$
and $\hat{p}_{+}=(\hat{p}_{A}+\hat{p}_{\mathrm{in}})/\sqrt{2}$ by
homodyne detection. So Alice sends the measurement outcomes, $m_{q}$
e $m_{p}$, through classical channels to Bob. In his turn, Bob performs
phase and amplitude modulations in his bipartite subsystem according
to the received classical information, that is, $\hat{q}_{\mathrm{out}}=\hat{q}_{B}-\sqrt{2}gm_{q}$
and $\hat{p}_{\mathrm{out}}=\hat{p}_{B}+\sqrt{2}gm_{p}$, in which
$g$ is the gain introduced in the classical communication or modulation.
We can notice that $g>1$ makes a teleportation with amplification,
whereas $g<1$ gives a teleportation with deamplification. Cases in
which the gain is different of unity are sometimes called teleamplifications
\cite{Neergaard-Nielsen13,He15}. To $g=1$ we get the usual protocol
proposed by Braunstein and Kimble. 

Taking into account the (de)amplification gain, the wanted states
by Bob must be $\{|\beta\rangle=|g\alpha\rangle\}_{B}$. However,
in a realistic situation, the actually transmitted states are described
by the set of density matrices $\{\rho_{\mathrm{out}}\}$ . So we
have to calculate the fidelity to the ideal task $|\alpha\rangle\rightarrow|\beta\rangle$
\cite{Jozsa94}, that is, 
\begin{equation}
F=\left[\mathrm{tr}\sqrt{\sqrt{\rho_{\beta}}\rho_{\mathrm{out}}\sqrt{\rho_{\beta}}}\right]^{2},\label{fidelity}
\end{equation}
where $\rho_{\beta}=|\beta\rangle\langle\beta|$. We can rewrite the
fidelity in terms of the respective Wigner functions, 
\begin{equation}
F=2\pi\int W_{\beta}(p,q)W_{\mathrm{out}}(p,q)dqdp,\label{fidelity-wigner}
\end{equation}
in which $W_{\beta}(p,q)$ and $W_{\mathrm{out}}(p,q)$ correspond
to $|\beta\rangle$ and $\rho_{\mathrm{out}}$, respectively. Proceeding
with the calculation of fidelity (\ref{fidelity-wigner}) as previous
references \cite{Chizhov02,Fiurasek02}, we obtain, then, 
\begin{equation}
F=\frac{2\exp\left[-\frac{1}{2}(\bm{x_{\beta}}-g\bm{x_{\alpha}})^{T}E_{t,g}^{-1}(\bm{x_{\beta}}-g\bm{x_{\alpha}})\right]}{\sqrt{\det(E_{t,g})}},\label{fide-out}
\end{equation}
with $\bm{x_{i}}=(2\Re(j)\;,\;2\Im(j))^{T}$, $j=\{\alpha,\beta\}$
and the matrix in denominator is 
\begin{equation}
E_{t,g}=(1+g^{2})D+g^{2}ZA_{t}Z^{T}-g(ZC_{t}+C_{t}^{T}Z^{T})+B_{t},\label{matrix-E}
\end{equation}
where $Z=\left(\begin{array}{cc}
1 & 0\\
0 & -1
\end{array}\right)$. For construction of the protocol, $\beta=g\alpha$, in the way that
it is trivial to calculate the mean fidelity, considering the prior
distribution (\ref{distribution-function}) of the input states. Therefore
\begin{equation}
\bar{F}(V_{t};g)=\frac{2}{\sqrt{\det(E_{t,g})}},\label{mean-fidelity}
\end{equation}
where the fidelity is explicitly described as function of the quantum
resources, represented by $V_{t}$, and the classical communication,
represented by $g$. This is a relevant result because all realistic
contributions, such as non-unity gain and losses of the entangled
system, are included in the term $\det(E_{t,g})$. So the complete
description of the teleportation protocol can be found thoroughly
analyzing it.

\section{Robust teleportation conditions}

The task to perform a non-unity gain teleportation, ideally represented
as $\{|\alpha\rangle\}_{A}\rightarrow\{|g\alpha\rangle\}_{B}$, has
a maximal fidelity for exclusively classical resources, using measure-and-prepare
strategies. As obtained in previous articles \cite{Namiki08,Chiribella13},
this classical fidelity threshold (CFT), using deterministic or probabilistic
processes, is 
\begin{equation}
F_{\mathrm{CFT}}(g)=\frac{1}{1+g^{2}},\label{fidelity-benchmark}
\end{equation}
that is, for a genuinely quantum teleportation, the average fidelity
must be strictly larger than $F_{\mathrm{CFT}}$.

Hence the genuinely quantum BK teleportation condition with attenuations
and non-unity gain is obtained comparing expressions (\ref{mean-fidelity})
and (\ref{fidelity-benchmark}), so that the classical regime is restricted
to $\bar{F}(V_{t,g})\leq F_{\mathrm{CFT}}(g)$, otherwise the process
is exclusively quantum. Such condition can be better analyzed if we
take only the determinant in (\ref{mean-fidelity}), so that the necessary
and sufficient condition of classical regime is restricted to $\det(E_{t,g})\geq4(1+g^{2})^{2}$.
This inequality depends on all entries of the bipartite system covariance
matrix, thus the obtained expressions are cumbersome to derive new
relations among the teleportation parameters, like $g$ and $t_{i}$.
On the other hand, we can search for symmetry properties of the average
fidelity $\bar{F}(V_{t,g})$, so that a quantity preserved by some
transformation could be found. In Appendix A, the following result
is proved: The average fidelity $\bar{F}(V_{t,g})$ is invariant under
local phase rotations of the shared bipartite system, $S_{\mathrm{inv}}:=R_{\theta A}\oplus R_{\theta B}\in SO(2,\Re)\oplus SO(2,\Re)$,
such that the rotation angles are constrained by $\theta_{B}=-\theta_{A}$,
in which indexes $A$ and $B$ label the modes shared by Alice and
Bob, respectively. 

With this invariance property of the fidelity, we can do a phase displacement
or a quadrature basis choice, to handle $\det(E_{t,g})$ without changing
the average fidelity, and, therefore, maintaining the relation of
the teleportation process with the CFT. Hence the classical-quantum
border can be calculated as presented in Appendix B, so that we obtain
the following result:

\textit{Result 1:} Given a BK teleportation with amplification gain
$g\geq0$, such that the shared bipartite system suffers local attenuations
$t_{A}$ and $t_{B}$, and choosing a quadrature basis according to
$S_{\mathrm{inv}}$, then the necessary and sufficient condition to
exist a classical measure-and-prepare strategy performing the same
task, i.e., to the fidelity be below the CFT, is 
\begin{eqnarray}
W_{\mathrm{all}}:=2(1+g^{2})\left[(gt_{A})^{2}(\mathrm{tr}(A)-2)+t_{B}^{\;2}(\mathrm{tr}(B)-2)-2gt_{A}t_{B}(K_{Q}-K_{P})\right] & +\nonumber \\
+\left[(gt_{A})^{2}(Q_{A}-1)+t_{B}^{\;2}(Q_{B}-1)-2gt_{A}t_{B}K_{Q}\right] & \times\label{quantum-condition}\\
\times\left[(gt_{A})^{2}(P_{A}-1)+t_{B}^{\;2}(P_{B}-1)+2gt_{A}t_{B}K_{P}\right] & \geq & 0.\nonumber 
\end{eqnarray}
Of course, this proposition has a converse, namely, the sufficient
and necessary condition to a genuinely quantum BK teleportation is
$W_{\mathrm{all}}<0$. Any way, this expression has many terms, but
we can rewrite it in more familiar ways. Consider the EPR-like operators
\begin{equation}
\hat{u}=gt_{A}\hat{q}_{A}-t_{B}\hat{q}_{B}\label{u-definition}
\end{equation}
and 
\begin{equation}
\hat{v}=gt_{A}\hat{p}_{A}+t_{B}\hat{p}_{B}.\label{v-definition}
\end{equation}
So condition (\ref{quantum-condition}) can be written as 
\begin{eqnarray}
\left[(2-t_{B}^{\;2})+\left(2-t_{A}^{\;2}\right)g^{2}\right] & \times\nonumber \\
\times\left\{ \langle(\Delta\hat{u})^{2}\rangle+\langle(\Delta\hat{v})^{2}\rangle-2\left[(gt_{A})^{2}+t_{B}^{\;2}\right]\right\}  & +\label{complete-condition}\\
+\langle(\Delta\hat{u})^{2}\rangle\langle(\Delta\hat{v})^{2}\rangle-\left[(gt_{A})^{2}+t_{B}^{\;2}\right]^{2} & \geq & 0,\nonumber 
\end{eqnarray}
where the variances are calculated to EPR-like operators (\ref{u-definition})
and (\ref{v-definition}).

As the variances are non-negative quantities, we can prove that the
third line in (\ref{complete-condition}) is negative, if the second
line in (\ref{complete-condition}) is also negative. Conversely,
the second line is positive, if the third line is positive as well
(see Barbosa et al. \cite{Barbosa11} for a similar deduction). In
addition, first line in (\ref{complete-condition}) is always positive.
Therefore we can split teleportation condition (\ref{complete-condition})
in two weaker conditions:

\textit{Result 2:} Given a BK teleportation with amplification gain
$g\geq0$, such that the shared bipartite system suffers local attenuations
$t_{A}$ and $t_{B}$, and choosing a quadrature basis according to
$S_{\mathrm{inv}}$, then a sufficient condition for its fidelity
is below the CFT is 
\begin{equation}
W_{\mathrm{prod}}:=\langle(\Delta\hat{u})^{2}\rangle\langle(\Delta\hat{v})^{2}\rangle-\left[(gt_{A})^{2}+t_{B}^{\;2}\right]^{2}\geq0.\label{necessary-condition}
\end{equation}

\textit{Result 3:} Given a BK teleportation with amplification gain
$g\geq0$, such that the shared bipartite system suffers local attenuations
$t_{A}$ and $t_{B}$, then a sufficient condition in order to surpass
the CFT, i.e., to be a genuinely quantum teleportation, is 
\begin{equation}
W_{\mathrm{sum}}:=\langle(\Delta\hat{u})^{2}\rangle+\langle(\Delta\hat{v})^{2}\rangle-2\left[(gt_{A})^{2}+t_{B}^{\;2}\right]<0.\label{sufficient-condition}
\end{equation}

Again, conditions (\ref{necessary-condition}) and (\ref{sufficient-condition})
have converses. In particular, from expression (\ref{necessary-condition}),
a necessary condition for a genuinely quantum teleportation is $W_{\mathrm{prod}}<0$.
It is very important to notice that Result 3 does not make reference
to basis choice. The reason is because $W_{\mathrm{sum}}$ is manifestly
invariant under the transformations $S_{\mathrm{inv}}$, in same way
that the average fidelity $\bar{F}(V_{t,g})$ is. Therefore condition
(\ref{sufficient-condition}) is as general as fidelity and CFT to
characterize the teleportation process. 

Conditions (\ref{necessary-condition}) and (\ref{sufficient-condition})
are more convenient than condition (\ref{complete-condition}) to
characterize a teleportation apparatus, since it is necessary less
knowledge about the shared bipartite system, that is, less covariance
matrix entries for measuring. On the other hand, we have to know the
gain $g$ and the attenuation transmissibilities $t_{A}$ and $t_{B}$.
Nevertheless realistic scenarios, which long-distance transmissions
of the correlated bipartite system would have unknown transmissibilities,
are reasonable and very probable in near future. Thus let us consider
an experimentalist could tune the gain of the teleportation amplification,
so that the process be maintained at a level above the CFT, independent
of any partial attenuation. Here it is worth to remark that the relevant
attenuations are only the partial ones, $t_{A};t_{B}>0$, because
in case of total attenuations, $t_{A}=0$ or $t_{B}=0$, the bipartite
system shared by Alice and Bob becomes separable, in fact, the partners
share no correlated system to perform teleportation, and there is
no quantum process. On the contrary, in the case of partial attenuations,
we have the following situations. First, one can call robust quantum
teleportation a process with tunable gain and which its shared bipartite
system is able to perform genuinely quantum teleportation for any
partial local attenuation. Second, if there are partial attenuations
that make the teleportation fidelity decrease below the CFT, the process
is called fragile teleportation. Third situation is that, for any
attenuation, the share bipartite system is unable to perform genuinely
quantum teleportation. Such situation always occurs to separable states
and to some possible entangled states.

So a relevant question for teleportation with unknown attenuations
is what Gaussian bipartite systems are able to perform robust quantum
teleportation. In order to find a condition that maximally delimitates
the bipartite system set able to robust quantum teleportation, we
have to minimize the expression $W_{\mathrm{sum}}$ to satisfy the
sufficient condition of genuinely quantum teleportation. It is clear
that $W_{\mathrm{sum}}$ as function of $g$ has a global minimum
at $g_{\mathrm{min}}=\frac{t_{B}(K_{Q}-K_{P})}{t_{A}(\mathrm{tr}(A)-2)}$.
Therefore, substituting $g_{\mathrm{min}}$ in expression (\ref{sufficient-condition}),
we obtain:

\textit{Result 4:} Given a BK teleportation, there is an amplification
gain, so that the sufficient condition to surpass the CFT, for any
local attenuations on the shared bipartite system, is 
\begin{equation}
W_{\mathrm{rob}}:=(\mathrm{tr}(A)-2)(\mathrm{tr}(B)-2)-(K_{Q}-K_{P})^{2}<0.\label{sufficient-robust-condition}
\end{equation}

Condition (\ref{sufficient-robust-condition}) is also manifestly
invariant under the transformations $S_{inv}$ (see Appendix B), so
the property of a bipartite system to be able to robust teleportation
follows the same generality of the characterization of a teleportation
with a determinate average fidelity. 

After these results, we may do some considerations. Firstly, the dependence
of the average fidelity of the relative phases of the input modes
has already been noted by Zhang et al. \cite{Zhang03}, in which an
explicit calculation has shown that the fidelity depends on the relative
phases among the shared bipartite modes and the signal teleported.
So phase fluctuations insert extra noise on the output signal, degrading
the average fidelity. Differently, in present article we derive the
transformations that retain the invariance of fidelity, which are
the phase changes with the constraint $\theta_{B}=-\theta_{A}$. Another
noteworthy aspect is the symmetry properties of Gaussian bipartite
systems obey a hierarchy. In terms of the phase space, the Gaussian
systems are preserved by general symplectic operations $Sp(4,\Re)$
\cite{Weedbrook12}; then the entanglement or separability feature
is preserved by the local symplectic operations $Sp(2,\Re)\oplus Sp(2,\Re)$
\cite{DGCZ00,Simon}; in addition robust bipartite entanglement is
maintained with local phase rotations $SO(2,\Re)\oplus SO(2,\Re)$,
as shown by Barbosa et al. \cite{Barbosa11}; and finally BK teleportation
fidelity is invariant under local rotation with angles $\theta_{B}=-\theta_{A}$.
In addition, we have also shown that the property of bipartite system
able to robust quantum teleportation are invariant to this last group,
the local bipartite rotations of anti-symmetric angles. Each presented
operation set forms a transformation group. These groups are related
by a subgroup chain, being the first the biggest group, following
until the smallest group. These sequence is associate to bipartite
system properties, following successively from the weakest to the
strongest property, respectively.

\section{Comparisons with other criteria}

Given the parameters $g$, $t_{A}$ and $t_{B}$ of the process, we
notice that the genuinely quantum teleportation condition from (\ref{sufficient-condition}),
$W_{\mathrm{prod}}<0$, is formally equivalent to a bipartite entanglement
condition obtained by Giovannetti et al. \cite{Giovannetti03} (see
also an early version by Tan \cite{Tan99}). However, in present article,
this condition is necessary for genuinely quantum teleportation, while
in the previous article \cite{Giovannetti03} the condition is sufficient
for entanglement. Considering the sufficient condition of genuinely
quantum teleportation (\ref{sufficient-condition}), $W_{\mathrm{sum}}<0$,
we also notice that it is formally equivalent to sufficient bipartite
entanglement conditions obtained by Simon \cite{Simon01} and Giovannetti
et al. \cite{Giovannetti03}. Considering these conditions and comparing
the Simon PPT condition \cite{Simon} with the complete condition
of genuinely quantum teleportation (\ref{complete-condition}), one
can verify that Gaussian bipartite entanglement and ability to perform
genuinely quantum teleportation are different properties. But, in
this comparison, we do not consider any operation of optimization
or adjustment of the protocol. 

On the other hand, it is interesting to rewrite $W_{\mathrm{sum}}$
in a more customary form. Discarding trivial cases, such as the case
$g=0$, that is, when there is no classical communication, and the
cases with $t_{A}=0$ or $t_{B}=0$, when the entangled beams are
completely attenuated, resulting in clearly classical cases, we define
the parameter 
\begin{equation}
\eta=\sqrt{\frac{gt_{A}}{t_{B}}}.\label{eta-definition}
\end{equation}
So the EPR-like operators from condition (\ref{sufficient-condition})
can be redefined as 
\begin{equation}
\check{u}=\eta\hat{q}_{A}-\frac{1}{\eta}\hat{q}_{B}\label{u-definition-other}
\end{equation}
and 
\begin{equation}
\check{v}=\eta\hat{p}_{A}+\frac{1}{\eta}\hat{p}_{B}.\label{v-definition-other}
\end{equation}
Hence condition (\ref{sufficient-condition}) becomes formally equivalent
to the sufficient entanglement condition of Duan et al. \cite{DGCZ00},
namely, 
\begin{equation}
\langle(\Delta\check{u})^{2}\rangle+\langle(\Delta\check{v})^{2}\rangle<2\left(\eta^{2}+\frac{1}{\eta^{2}}\right).\label{sufficient-condition-other}
\end{equation}
 This condition is a usual criterion to test the entanglement of continuous-variable
bipartite systems. Its popularity is devoted to few necessary terms
to measure, only two variances. Thus we can consider that expression
(\ref{sufficient-condition-other}) is also a test valid to verify
if a bipartite system is useful to teleportation task. 

Result 4 presented in this paper is also very connected to an entanglement
condition. In fact, condition (\ref{sufficient-robust-condition})
is very close to a previous result due to Barbosa et al. \cite{Barbosa11}:

\textit{Result from \cite{Barbosa11}:} Given an initially entangled
bipartite system, its entanglement is robust to any partial local
attenuations if only if 
\begin{equation}
W_{\mathrm{full}}:=(\mathrm{tr}(A)-2)(\mathrm{tr}(B)-2)-\mathrm{tr}(C^{T}C)+2\det(C)\leq0,\label{barbosa}
\end{equation}
for Gaussian states. Condition (\ref{barbosa}) is only sufficient
for non-Gaussian states.

With a suitable choice of quadrature basis, according to $S_{\mathrm{inv}}$,
inequality (\ref{barbosa}) reduces to condition $W_{\mathrm{rob}}\leq0$.
Hence almost all robust Gaussian bipartite states are also Gaussian
bipartite states able to robust quantum teleportation. With exception
of a very small set of borderline states, namely, entangled states
with $W_{\mathrm{full}}=0$, we have $W_{\mathrm{full}}<0$ and $W_{\mathrm{rob}}<0$
delimitate the same state set. Therefore we claim that the robust
Gaussian bipartite entanglement is essentially equivalent to Gaussian
bipartite system property of being able to robust quantum teleportation.
We notice that the induction of the robustness property from entanglement
to teleportation is also observed in article of Adesso et al. \cite{Adesso05},
where the equivalence between Gaussian bipartite entanglement and
genuinely quantum teleportation, whose fidelity is optimized. Such
parallelism stresses the relevance of the robustness in entangled
systems.

\section{Some examples}

To clarify the relations among the teleportation conditions found
in this article and the previous entanglement conditions, now we consider
some specific situations in what follows. A way to verify these relationships
is plotting the regions of physically possible states of the shared
bipartite system as function of relevant parameters. Considering that
the shared bipartite system used in teleportation is described by
following symmetric covariance matrix, 
\begin{equation}
V=\left(\begin{array}{cccc}
Q & 0 & K_{Q} & 0\\
0 & P & 0 & K_{P}\\
K_{Q} & 0 & Q & 0\\
0 & K_{P} & 0 & P
\end{array}\right),\label{symmetric-covariance}
\end{equation}
we can plot Figure 1 as a function of $\overline{K_{Q}}:=K_{Q}/Q$
and $\overline{K_{P}}:=K_{P}/P$. The colorful and numbered regions
indicate states with different correlation properties. In both Figures
1a and 1b, share bipartite system states able to teleportation and
robust to any partial local attenuation ($0<t_{i}\leq1$, $i=A;B$)
lie in region I (red). States able to quantum teleportation, but fragile
to some partial local attenuation, are presented in region II (light
blue in Figure (a) and light purple in Figure (b)). Separable states
are comprised within region IV (yellow). Entangled states which are
not able to genuine quantum teleportation (with fidelity below the
CFT) lie in regions III (blue and light green in Figure (b)) and V
in Figure (b) (purple and orange). The white regions in both Figures
1a and 1b are unphysical states, prohibited by the Robertson\textendash{}Schrödinger
uncertainty relation and the purity \cite{Simon94,Adesso04}, $0<(\det(V))^{-2}\leq1$.
The plots are limited to $|\overline{K_{Q}}|;|\overline{K_{P}}|\leq1$,
due to the Schwartz inequality. Figure I presents two situations for
comparison of a symmetrized/optimized teleportation (Figure 1a) and
a non-optimized teleportation (Figure 1b). In the Figure 1a, the attenuations
and gain are such that $\eta=gt_{A}/t_{B}=1$. So in this case, the
border between regions I and II is given by coincident conditions
(\ref{sufficient-condition}) and (\ref{sufficient-robust-condition}),
(the border on $W_{\mathrm{sum}}=0$ and $W_{\mathrm{rob}}=W_{\mathrm{full}}=0$);
regions II and III are delimited by condition (\ref{complete-condition})
(border on $W_{\mathrm{all}}=0$), and the border between regions
III and IV is determined by condition (\ref{necessary-condition})
(border on $W_{\mathrm{prod}}=0$) and the Simon PPT condition, which
are coincident in this particular case. However, the situation is
more complex in Figure 1b, where it is taken $\eta=gt_{A}/t_{B}=0.65$
and $t_{B}=1$. Here condition to $W_{\mathrm{sum}}=0$ is displaced
to delimit regions I and II, and condition $W_{\mathrm{rob}}=W_{\mathrm{full}}=0$
is maintained, delimiting regions III and V, because it does not depend
on the parameters $g$, $t_{A}$ and $t_{B}$. Condition $W_{\mathrm{all}}=0$
is also displaced to delimit regions II and V. At last, condition
(\ref{necessary-condition}) and the Simon PPT condition are no longer
coincident. PPT condition stays set apart region IV from the other
states, but condition $W_{\mathrm{prod}}=0$ is displaced to separate
regions (III/V) a from b. We can notice that regions useful to teleportation
are reduced if $g$ is not optimized. However, condition (\ref{sufficient-robust-condition})
establishes what shared bipartite states can become able to robust
teleportation, given a suitable gain, extending red region I in Figure
1b to regions II and V. Therefore the robust teleportation state regions
are the same in Figures 1a and 1b.

\begin{figure}[H]
\noindent \begin{centering}
\includegraphics[width=16cm]{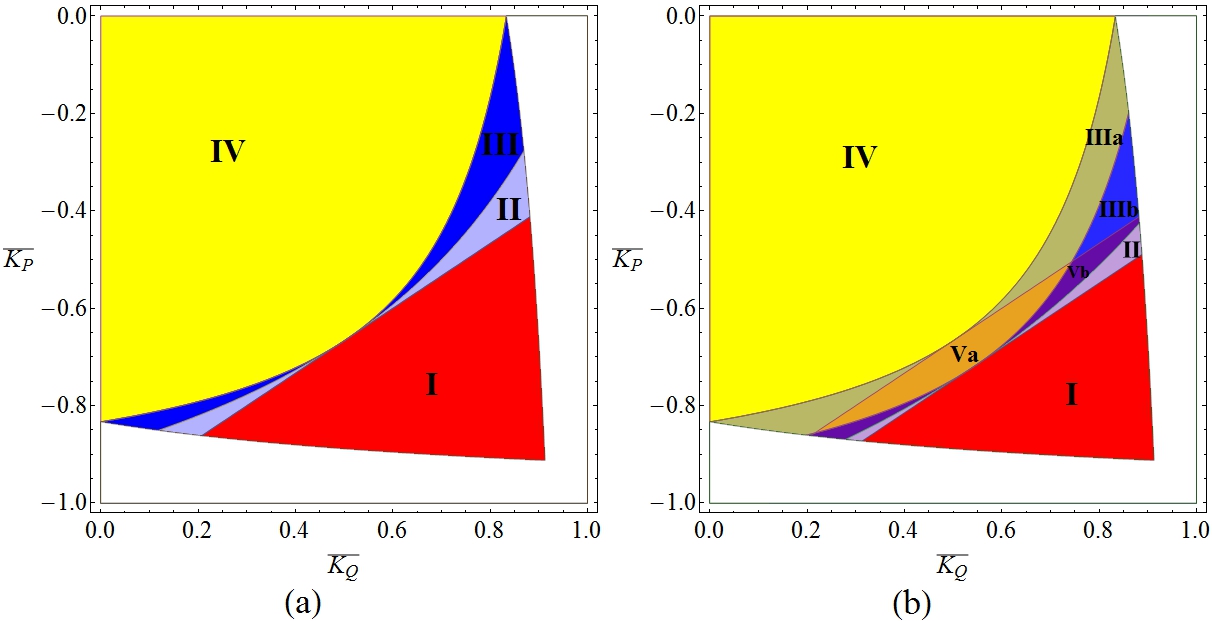}
\par\end{centering}

\noindent \centering{}\caption{The space of states of the Gaussian bipartite system with covariance
matrix (\ref{symmetric-covariance}) is presented as a function of
$\overline{K_{Q}}:=K_{Q}/Q$ and $\overline{K_{P}}:=K_{P}/P$. In
Figure (a), the teleportation is performed tuning gain as $gt_{A}/t_{B}=1$.
Bipartite system states able to robust quantum teleportation lie in
region I (red). States able to quantum teleportation, but fragile
to some partial local attenuation, are presented in region II (light
blue). Entangled states which are not able to genuine quantum teleportation
(with fidelity below the CFT) lie in region III (blue). Separable
states are comprised within region IV (yellow). In Figure (b), the
teleportation is performed with $t_{B}=1$ and $gt_{A}/t_{B}=0.65$.
As in Figure (a), robust quantum teleportation states lie in region
I (red), fragile quantum teleportation states lie in region II (light
purple), and separable states lie in region IV (yellow). Moreover,
as $g$ is not optimized, regions IIIa/b and Va/b represent the entangled
states unable to genuine quantum teleportation. However, the gain
$g$ can always be tuned, so that region I is maximized, expanding
on regions II and V. In both Figures, the white region represents
unphysical states.}
\end{figure}

In addition, we can study specific states of the shared bipartite
system, to visualize the dynamics of the average fidelity in terms
of the local attenuations and the gain. In Figure 2, it is plotted
the teleportation fidelity as function of the attenuation transmissibilities
$t_{A}$ and $t_{B}$, for unity gain and the shared bipartite system
being initially in a symmetric two-mode squeezed state, such that
the covariance matrix entries are $Q_{A}=P_{A}=Q_{B}=P_{B}=\cosh(2r)$,
$K_{Q}=-K_{P}=\sinh(2r)$, and null in other cases. In Figure 2, the
CFT is represented by the thick straight lines. The clearer coned
region is where the teleportation surpasses the CFT. A very similar
plot was obtained by He et al. \cite{He15}. In that study, however,
the delimited coned region is concerning the so-called secure teleportation,
that is, teleportation with the fidelity above the telecloning threshold.
Hence in that paper the mentioned region is narrower than the present
study, because the telecloning threshold is 2/3, for unity gain. This
dynamic of the fidelity in terms of the transmissibilities is persistent
for symmetric bipartite states. But as we shall see, this is not true,
when considering asymmetric states.

\begin{figure}[H]
\noindent \begin{centering}
\includegraphics[width=8.5cm]{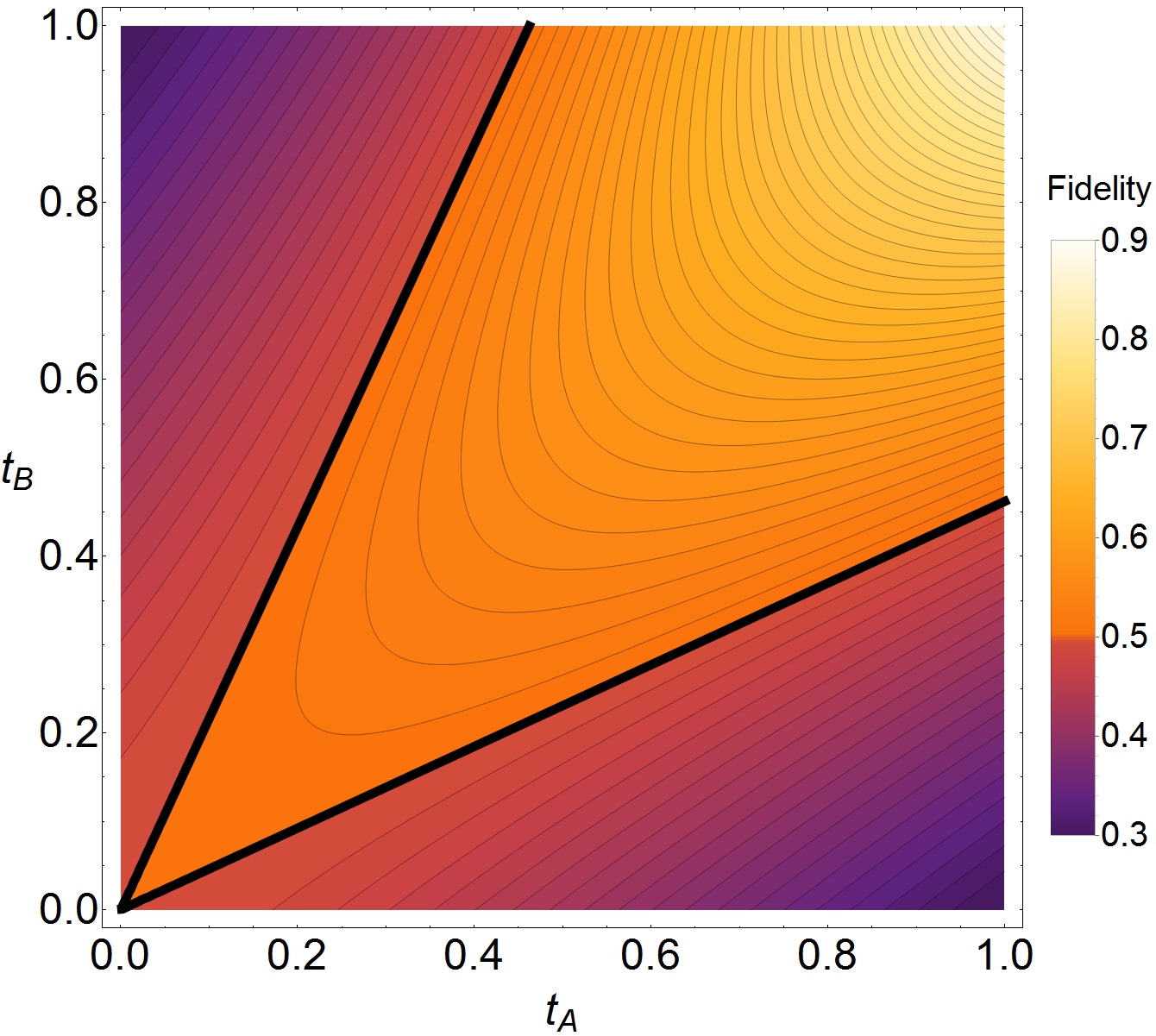}
\par\end{centering}

\caption{Fidelity as a function of the transmissibilities $t_{A}$ and $t_{B}$.
The quantum teleportation is possible inside the coned region. In
this plot, the gain is $g=1$ and the entangled bipartite system is
a symmetric two-mode squeezed state with squeezing parameter $r=1$.}
\end{figure}

With a direct verification of condition (\ref{sufficient-robust-condition}),
the state considered in Figure 2 is able to robust quantum teleportation,
although there are non-null values of $t_{A}$ and $t_{B}$ so that
the fidelity is below the CFT. However, this case is fixed to $g=1$,
and the definition of robust quantum teleportation is that the teleportation
process, whose amplification gain can be tuned, has an average fidelity
larger than the CFT for any local partial attenuation. Hence the coned
region in Figure 2 can be displaced to cover all $t_{A}\times t_{B}$
plane, as the gain is tuned to this purpose. This is observed in Figure
3, in which is used the same original state of Figure 2. In Figure
3a, the gain is $g=0.5$, so the coned region is shifted toward the
axis $t_{A}$. As $g\rightarrow0$, the coned region continuously
approaches to this axis. A side effect is that the average fidelity
increases, as well as the CFT. Opposite effects are found in the case
of gain $g=2.5$, in Figure 2b, in which the coned region is shifted
toward the axis $t_{B}$. As $g\rightarrow\infty$, the coned region
continuously approaches to this axis, and the average fidelity and
the CFT decrease. These characteristics are found in all sufficiently
symmetrical states, which for many practical purposes are interesting
and useful for communication and processing of the quantum information. 

\begin{figure}[H]
\noindent \begin{centering}
\includegraphics[width=16cm]{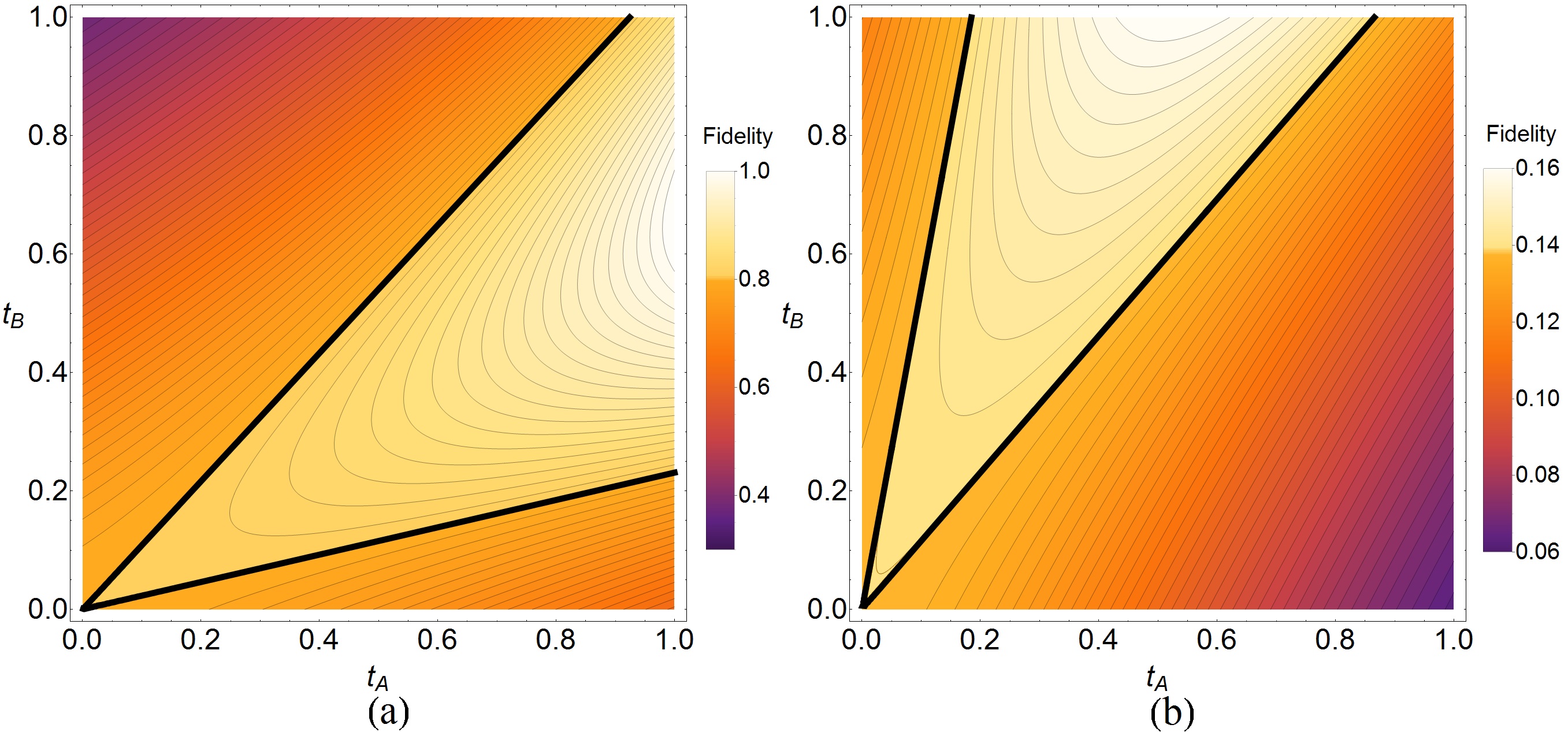}
\par\end{centering}

\caption{Fidelity as a function of the transmissibilities $t_{A}$ and $t_{B}$.
The quantum teleportation is possible inside the coned regions. In
these plots, the entangled bipartite system is same as in Figure 2.
In Figure (a), the gain is $g=0.5$. In Figure (b), the gain is $g=2.5$.
We notice that this bipartite system is able to robust quantum teleportation,
because the coned region sweeps all plane $t_{A}\times t_{B}$, except
regions of total attenuation ($t_{i}=0$), as the gain is tuned, inside
the range $g\in(0,\infty)$. }
\end{figure}

Beyond the symmetric states, it is also important to fully characterize
all possible scenarios for bipartite generic Gaussian systems, because
in further applications of the teleportation, the modes sent to Alice
and Bob could be ill generated and undergo unknown attenuations. Therefore
we have to focus the asymmetric bipartite states as well. Considering
a case such that 
\begin{equation}
V=\left(\begin{array}{cccc}
2.1 & 0 & 1.9 & 0\\
0 & 2.6 & 0 & -0.7\\
1.9 & 0 & 2.2 & 0\\
0 & -0.7 & 0 & 2.4
\end{array}\right),\label{asymmetric-covariance}
\end{equation}
and with unity gain, one can plot Figure 4, in which the trick curve
assigns the CFT. We notice that the genuinely quantum teleportation
is only maintained for transmissibility values close to 1, observed
in the region surrounded in the upper right corner of Figure 4. For
non-unity gain values, the genuinely quantum teleportation region
tends to decrease, so that the quantum region cannot cover all $t_{A}\times t_{B}$
plane, even tuning the gain. So this is a typical case of fragile
bipartite states for quantum teleportation.

\begin{figure}[H]
\noindent \begin{centering}
\includegraphics[width=8.5cm]{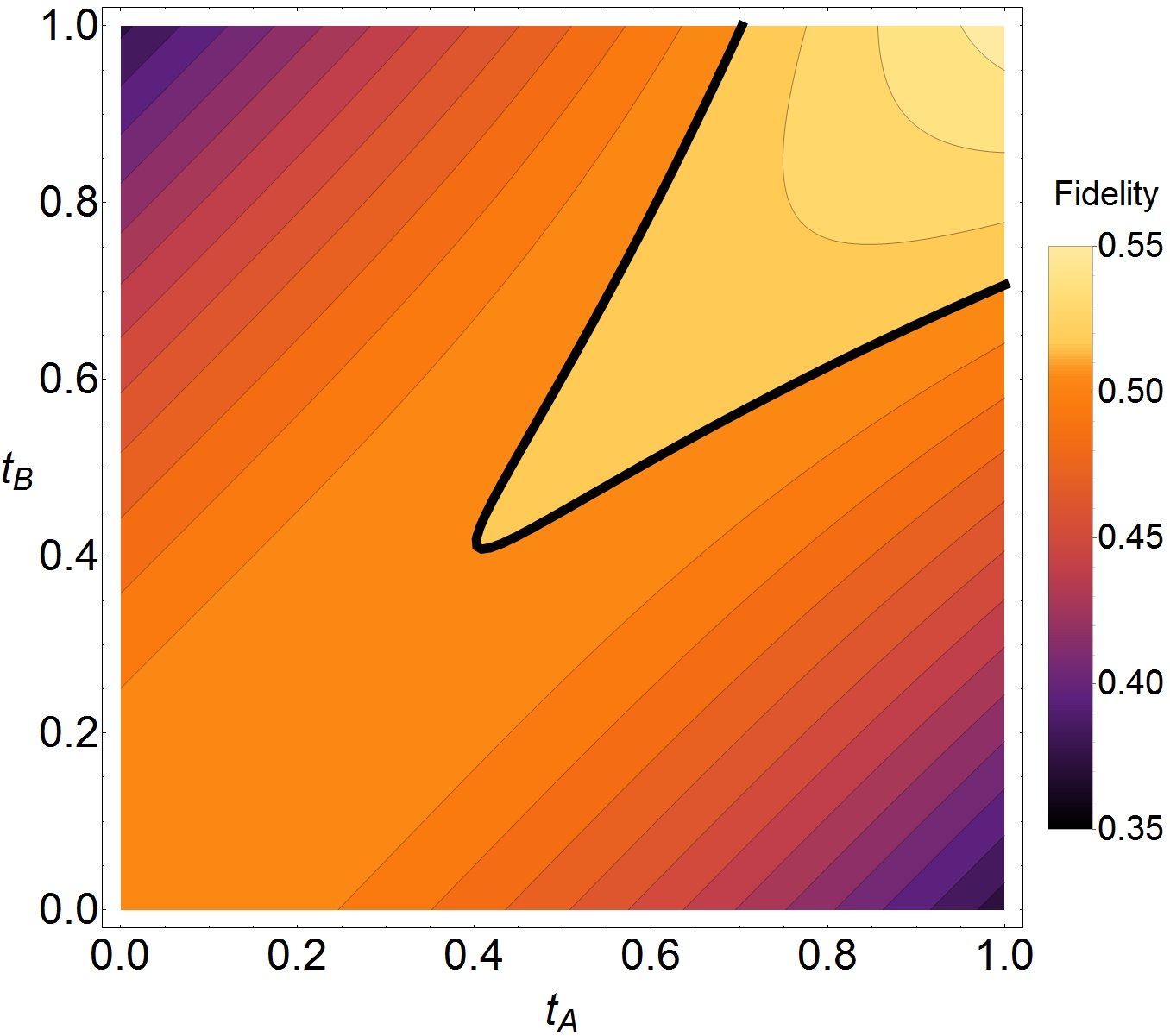}
\par\end{centering}

\noindent \centering{}\caption{Fidelity as a function of the transmissibilities $t_{A}$ and $t_{B}$.
The quantum teleportation is possible inside the upper right region.
In this plot, the gain is $g=1$ and the entangled bipartite system
is an asymmetric two-mode state with covariance matrix given by Eq.
(\ref{asymmetric-covariance}). This example presents a fragile bipartite
system to perform teleportation, that is, even tuning the gain, the
quantum teleportation region does not cover all values of $0<t_{i}\leq1$.}
\end{figure}

Finally, there are entangled bipartite states that, together the separable
states, is not useful for a teleportation with fidelity above the
CFT. Once these cases are trivial, a respective figure is not presented.

\section{Discussion}

In this article, we have studied the details of the relationship between
the gain of the classical communication and the losses of the correlated
bipartite system shared by Alice and Bob, in the teleportation of
coherent state signals. We have found several conditions that characterize
the BK teleportation as functions of the its parameters, namely, the
bipartite system covariance matrix, the gain, and the attenuation
transmissibilities. These conditions provide criteria to determinate
if the teleportation process is successful for its fidelity being
above the classical threshold (CFT) of measure-and-prepare methods.
From that, given a suitable gain tuning, it was obtained a condition
to the shared bipartite system to be able to execute BK teleportation
robust to local partial attenuations. Such condition to robust quantum
teleportation was verified to be essentially equivalent to the robust
bipartite entanglement condition, showing that the robustness of the
entanglement is induced in the teleportation task. Along the presented
derivations, fidelity symmetry properties were found, revealing that
the main results obey the same transformation invariance of the average
fidelity. It is expected that these findings are helpful to future
long-distance teleportation implementations, in which one wants to
generate a shared bipartite system useful for this task, independently
of unknown attenuations. 

As an initial motivation was to study the realistic conditions to
perform the teleportation, this article has a natural path to be generalized,
considering other dissipative dynamics of the quantum channels of
the bipartite system, for example, phase losses or atmospheric turbulence,
as it was studied by Bohmann et al. \cite{Bohmann16}. Other possible
extension of the presented studies treats about the teleported signal.
We have studied the case of transmitted signals belonging to an uniform
distribution of coherent states. However we can consider other sets
of coherent states, so that the teleportation task is more challenging
and diverse, including more complex dynamics combined with quantum
channel noise \cite{Haseler10}. Following in this way, future progresses
can generalize the present article to may other states of the input
signal, for example, general Gaussian pure states, in particular squeezed
states, since the respective CFTs have already been obtained by Chiribella
and Adesso \cite{Chiribella14}, or qbits and single photons \cite{Ide01},
whose implementations have been accomplished in recent years \cite{Takeda13a,Takeda13b,Kogias14}.
At last, further connections between the robust entanglement and teleportation
conditions and other quantum properties, like EPR correlations, will
possibly be established, as well as the secure teleportation is connected
to EPR steering \cite{He15}.
\begin{acknowledgments}
The author thanks Marcelo Martinelli for introducing the issue of
the non-unity gain teleportation and for enlightening discussions.
This research was supported by FAPEMIG (Fundação de Amparo à Pesquisa
do Estado de Minas Gerais), Grant No. CEX-APQ-01899-13.
\end{acknowledgments}
\appendix

\section{Symmetry Transformations of Fidelity}

The average fidelity of the teleportation of coherent states with
classical channel gain $g\geq0$ and independent attenuations of the
bipartite system modes is given by $\bar{F}(V_{t};g)=\frac{2}{\sqrt{\det(E_{t,g})}}$
(see, e.g., \cite{Fiurasek02}). Regardless of symmetry considerations,
the determinant is 
\begin{eqnarray}
 &  & \det(E_{t,g})=4(1+g^{2})^{2}+\nonumber \\
 &  & +2(1+g^{2})\left[(gt_{A})^{2}(Q_{A}+P_{A}-2)+t_{B}^{\;2}(Q_{B}+P_{B}-2)-2gt_{A}t_{B}(K_{Q}-K_{P})\right]+\nonumber \\
 &  & +\left[(gt_{A})^{2}(Q_{A}-1)+t_{B}^{\;2}(Q_{B}-1)-2gt_{A}t_{B}K_{Q}\right]\times\label{E-determinant-App}\\
 &  & \times\left[(gt_{A})^{2}(P_{A}-1)+t_{B}^{\;2}(P_{B}-1)+2gt_{A}t_{B}K_{P}\right]+\nonumber \\
 &  & -\left[(gt_{A})^{2}K_{A}-t_{B}^{\;2}K_{B}+2gt_{A}t_{B}(K_{1}-K_{2})\right]^{2}.\nonumber 
\end{eqnarray}

Managing this expression is difficult, particularly the correlations
in the last line of (\ref{E-determinant-App}). However cumbersome
derivations can be bypassed using symmetry transformations which preserve
$\bar{F}(V_{t,g})$ invariant. As we are studying the shared bipartite
system properties in Gaussian teleportation, the transformations must
preserve the Gaussian feature of the system. So the transformations
belong to unitary linear Bogoliubov maps \cite{Weedbrook12}. Furthermore
we must seek operations that, at least, have the basic property of
preserving entanglement, which is reserved for the local maps. These
transformations correspond to affine symplectic maps acting on the
phase space. To covariance matrices, these maps operate as $V\rightarrow SVS^{T}$
, where $S$ is a $2N\times2N$ real symplectic matrix, so that $N$
is the number of subsystems. Some of its properties are $S\Omega S^{T}=\Omega$,
such that $[\hat{\bm{x}},\hat{\bm{x}}^{T}]=2i\Omega$, and $\det(S)=1$.
In the case of a bipartite system shared by Alice and Bob, the general
symplectic transformations are $S_{AB}\in Sp(4,\Re)$, using the index
A to Alice and B to Bob. Then we must seek a symplectic operation,
$S_{\mathrm{inv}}$, on the covariance matrix of the shared bipartite
system, such that 
\begin{equation}
\bar{F}(V_{t};g)=\bar{F}(S_{\mathrm{inv}}V_{t}S_{\mathrm{inv}}^{\phantom{}T};g).\label{fidelity-transform}
\end{equation}
As the local transformations are a subgroup of $Sp(4,\Re)$, that
is, $S_{\mathrm{local}}=S_{A}\oplus S_{B}\in Sp(2,\Re)\oplus Sp(2,\Re)$.
For each single subsystem, the local symplectic transformations $S_{i}\in Sp(2,\Re)$,
$i=A;B$, can always be written as a product of three special maps,
\begin{equation}
S_{i}=R_{\theta}Y_{r}R_{\phi},\label{general-symplectic}
\end{equation}
where $Y_{r}$ is a squeezing operation and $R_{\theta}$ and $R_{\phi}$
are space-phase rotations. 

Moreover, $S_{\mathrm{inv}}$ must be independent of the quantum channels,
through which the bipartite system is delivered to Alice and Bob,
so 
\begin{equation}
S_{\mathrm{inv}}V_{t}S_{\mathrm{inv}}^{\phantom{}T}=S_{\mathrm{inv}}\mathcal{L}(V)S_{\mathrm{inv}}^{\phantom{}T}=\mathcal{L}(S_{\mathrm{inv}}VS_{\mathrm{inv}}^{\phantom{}T}).\label{covariance-transform}
\end{equation}
 As it is shown in the paper by Barbosa et al. \cite{Barbosa11},
in the case of Gaussian attenuations, equation (\ref{covariance-transform})
is valid only if $S_{i}S_{i}^{T}=I$, to each mode $i$. This condition
limits $S_{\mathrm{inv}}$ to the phase-space local rotation group,
$S_{\mathrm{inv}}=R_{A}\oplus R_{B}\in SO(2,\Re)\oplus SO(2,\Re)$.
For each mode, the rotations have the property of $RZ=ZR^{T}$, so
we obtain 
\begin{eqnarray}
E_{t,g} & = & (1+g^{2})D+g^{2}ZR_{A}A_{t}R_{A}^{T}Z^{T}-g(ZR_{A}C_{t}R_{B}^{T}+R_{B}C_{t}^{T}R_{A}^{T}Z^{T})+R_{B}B_{t}R_{B}^{T}=\nonumber \\
 & = & (1+g^{2})D+R_{A}^{T}(g^{2}ZA_{t}Z^{T})R_{A}-R_{A}^{T}gZC_{t}R_{B}^{T}-R_{B}gC_{t}^{T}Z^{T}R_{A}+R_{B}B_{t}R_{B}^{T}=\nonumber \\
 & = & R_{\theta}[(1+g^{2})D+g^{2}ZA_{t}Z^{T}-g(ZC_{t}+C_{t}^{T}Z^{T})+B_{t}]R_{\theta}^{T},\label{arg-transform}
\end{eqnarray}
where $R_{B}=R_{A}^{T}=R_{A}^{-1}=R_{\theta}$ and $\det(R_{\theta})=1$.
Therefore the fidelity is invariant under local phase rotations of
the shared bipartite system, constrained to $\theta_{B}=-\theta_{A}=\theta$,
namely, 
\begin{equation}
S_{\mathrm{inv}}\in\{R_{A}(\theta_{A})\oplus R_{B}(\theta_{B})\in SO(2,\Re)\oplus SO(2,\Re)|\theta_{A}=-\theta_{B}\}.\label{transform-invariance}
\end{equation}

\section{Properties of $\det(E_{t,g})$ under $S_{\mathrm{inv}}$}

Since we have established the mapping on $V_{t}$ that keeps invariant
$\bar{F}(V_{t};g)$, we can show equation (\ref{E-determinant-App})
has terms which are not relevant to $\bar{F}(V_{t};g)$, given a suitable
choice of the shared system quadrature basis. We will show that there
is always an $S_{\mathrm{inv}}$, such that $V\mapsto V^{\prime}$
and 
\begin{equation}
\left[(gt_{A})^{2}K_{A}-t_{B}^{\;2}K_{B}+2gt_{A}t_{B}(K_{1}-K_{2})\right]\mapsto\left[(gt_{A})^{2}K_{A}^{\prime}-t_{B}^{\;2}K_{B}^{\prime}+2gt_{A}t_{B}(K_{1}^{\prime}-K_{2}^{\prime})\right]=0.\label{basis-transform}
\end{equation}
To obtain this, we explicitly calculate the prime entries in expression
(\ref{basis-transform}), under the transformation $S_{\mathrm{inv}}=\left(\begin{array}{cc}
\cos\theta_{A} & \sin\theta_{A}\\
-\sin\theta_{A} & \cos\theta_{A}
\end{array}\right)\oplus\left(\begin{array}{cc}
\cos\theta_{B} & \sin\theta_{B}\\
-\sin\theta_{B} & \cos\theta_{B}
\end{array}\right)$. So 
\begin{eqnarray}
 &  & (gt_{A})^{2}\left[\frac{1}{2}(Q_{A}-P_{A})\sin(2\theta_{A})+K_{A}\cos(2\theta_{A})\right]+\nonumber \\
 &  & -t_{B}^{\;2}\left[\frac{1}{2}(Q_{B}-P_{B})\sin(2\theta_{B})+K_{B}\cos(2\theta_{B})\right]+\label{transformed-entries}\\
 &  & +2gt_{A}t_{B}\left[(K_{Q}+K_{P})(\cos\theta_{A}\sin\theta_{B}-\sin\theta_{A}\cos\theta_{B})+\right.\nonumber \\
 &  & \left.+(K_{1}-K_{2})(\cos\theta_{A}\cos\theta_{B}+\sin\theta_{A}\sin\theta_{B})\right]=0.\nonumber 
\end{eqnarray}
To $S_{\mathrm{inv}}$, we must have $\theta_{A}=-\theta_{B}=\theta$.
Collecting $\theta$, we get 
\begin{equation}
\theta=\frac{1}{2}\arctan\left[\frac{(gt_{A})^{2}K_{A}-t_{B}^{\;2}K_{B}+2gt_{A}t_{B}(K_{1}-K_{2})}{\frac{1}{2}(gt_{A})^{2}(Q_{A}-P_{A})+\frac{1}{2}t_{B}^{\;2}(Q_{B}-P_{B})-2gt_{A}t_{B}(K_{Q}+K_{P})}\right].\label{transform-angle}
\end{equation}
As the arctan domain is all $\Re$ and arctan is a many-valued function,
so there are always multiple values to $\theta$. Therefore, we can
always write the covariance matrix in a phase-space basis, so that
$\left[(gt_{A})^{2}K_{A}-t_{B}^{\;2}K_{B}+2gt_{A}t_{B}(K_{1}-K_{2})\right]=0$,
keeping $\bar{F}(V_{t};g)$ invariant. Using such phase-space basis,
the condition (\ref{quantum-condition}) can be established.

Another important property of $V$ under transformation $S_{\mathrm{inv}}$
is that conditions (\ref{sufficient-condition}) and (\ref{sufficient-robust-condition})
are also manifestly invariant. To show this, we must notice that $\mathrm{tr}(A)$,
$\mathrm{tr}(B)$ and $(K_{Q}-K_{P})$ are invariant for any $R_{A}$
and $R_{B}$. As we can write $W_{\mathrm{sum}}=(gt_{A})^{2}(\mathrm{tr}(A)-2)+t_{B}^{\;2}(\mathrm{tr}(B)-2)-2gt_{A}t_{B}(K_{Q}-K_{P})$,
its invariance is clear. The same is valid to $W_{\mathrm{rob}}$.

\end{document}